\documentclass[
 aps,
 prd,
 twocolumn,
 showpacs,
 amsmath,
 amssymb,
 superscriptaddress
 ]{revtex4}

 \usepackage{graphicx}  
 \usepackage{dcolumn}   
 \usepackage{url}       

\def\u#1{${\mathbf{#1}}$}

\begin{document}


\title{How to Sum Contributions into the Total \\
       Charged-Current Neutrino--Nucleon Cross Section}
\affiliation{Bogoliubov Laboratory of Theoretical Physics,
             Joint Institute for Nuclear Research, RU-141980 Dubna, Russia}
\affiliation{Dzhelepov Laboratory of Nuclear Problems,
             Joint Institute for Nuclear Research, RU-141980 Dubna, Russia}
\affiliation{Institute for Theoretical and Experimental Physics, 
             RU-117259 Moscow, Russia}
\affiliation{Physics Department of Irkutsk State University,
             RU-664003 Irkutsk, Russia}
\affiliation{INFN, Sezione di Firenze, I-50019 Sesto Fiorentino (FI), Italy}
\author{Konstantin S. Kuzmin}
 \email{kkuzmin@theor.jinr.ru}
 \affiliation{Bogoliubov Laboratory of Theoretical Physics,
             Joint Institute for Nuclear Research, RU-141980 Dubna, Russia}
 \affiliation{Institute for Theoretical and Experimental Physics, 
              RU-117259 Moscow, Russia}
\author{Vladimir V. Lyubushkin}
 \email{lvv@nusun.jinr.ru}
 \affiliation{Dzhelepov Laboratory of Nuclear Problems,
             Joint Institute for Nuclear Research, RU-141980 Dubna, Russia}
 \affiliation{Physics Department of Irkutsk State University,
              RU-664003 Irkutsk, Russia}
\author{Vadim A. Naumov}
 \email{vnaumov@theor.jinr.ru}
 \homepage{http://theor.jinr.ru/~vnaumov/}
 \affiliation{Bogoliubov Laboratory of Theoretical Physics,
             Joint Institute for Nuclear Research, RU-141980 Dubna, Russia}
 \affiliation{INFN, Sezione di Firenze, I-50019 Sesto Fiorentino (FI), Italy}

\date{\today} 

\begin{abstract}
The total CC (anti)neutrino-nucleon cross section is usually estimated by
the sum of contributions from quasi-elastic scattering (QES), single-pion
production through baryon resonances (RES), and deep inelastic scattering (DIS)
with an appropriate scratching the phase space of the RES and DIS contributions.
However the resulting total cross section is very sensitive to the value of
the cut-off in invariant mass of the final hadron system produced in RES and DIS.
We examine available experimental data on the QES and total CC cross sections
in order to extract the best-fit value for this cut-off. By using the same data 
set we attempt to adjust the poorly known values of the axial mass for QES and RES.
\end{abstract}

  \pacs{12.15.Ji, 13.15.+g, 14.20.Gk, 23.40.Bw, 25.30.Pt}

  \keywords{Neutrino--nucleon interactions;
            Charged currents;
            Axial-vector form factors;
            Baryon resonances;
            Likelihood analysis   
           }

\maketitle

\section{Introduction}
  \label{Introduction}

It is conventional to estimate the inclusive charged and neutral current
neutrino--nucleon cross sections by the sum of contributions from
exclusive channels and deep inelastic scattering (DIS):
\begin{equation}\label{SymbolicSum}
 \sigma_{{\nu}N}^{\text{tot}}=
 \sigma_{{\nu}N}^{\text{(Q)ES}} \oplus
 \sigma_{{\nu}N}^{1\pi} \oplus
 \sigma_{{\nu}N}^{2\pi} \oplus\ldots\oplus
 \sigma_{{\nu}N}^{1K} \oplus\ldots\oplus
 \sigma_{{\nu}N}^{\text{DIS}}.
\end{equation}
In the absence of a received model for multi-hadron exclusive neutrinoproduction,
the exclusive contributions in Eq.~\eqref{SymbolicSum} are usually assumed to be
saturated by elastic (NC case) or quasielastic (CC case) scattering (ES/QES) and
single-pion production through baryon resonances (RES).
The exclusive and inclusive (DIS) contributions are of the same order of magnitude
within the few-GeV energy region. Thus, to avoid double counting, the phase space
of the RES and DIS contributions have to be scratched by the conditions
$W<W_{\text{cut}}^{\text{RES}}$ and $W>W_{\text{cut}}^{\text{DIS}}$, respectively,
where $W$ is the invariant mass of the final hadron system in RES or DIS, and
$W_{\text{cut}}^{\text{RES}}$ and $W_{\text{cut}}^{\text{DIS}}$ are some parameters.

The physical basis of this approximation is the concept of quark-hadron duality,
according to which the resulting total cross section should be essentially
independent of the specific values of the cutoff parameters if they are of the
order of the threshold value of $W$ for two-pion production,
$W_{2\pi}=M_N+2m_\pi\approx 1.2~\text{GeV}$. In practice, this value is too small
since the structure functions involved into the calculations of the DIS cross
section cannot be extrapolated to the two-pion production threshold due to the
obvious reasons.

The problem is aggravated by the uncertainties in the knowledge of the simplest
exclusive contributions: the description of the RES reactions is vastly model-dependent
and, even within a fixed model for RES, both RES and (Q)ES cross sections are very
sensitive to the poorly known shape of the weak axial-vector form factors. By adopting
the standard dipole parametrization for these form factors, their shapes can be described
with the two phenomenological parameters (``axial masses'') $M_A^{\text{QES}}$ and
$M_A^{\text{RES}}$ which, strictly speaking, may be different and whose experimental
values spread within inadmissibly wide ranges~\cite{Bernard:01}.

In this study we attempt to fine-tune both the axial masses $M_A^{\text{QES}}$,
$M_A^{\text{RES}}$ and the cutoffs $W_{\text{cut}}^{\text{RES}}$,
$W_{\text{cut}}^{\text{DIS}}$ by fitting available experimental
data on the QES with $\Delta Y=0$ and total CC cross sections for ${\nu}_\mu$ and
$\overline{\nu}_\mu$ scattering off different nuclear targets (converted to the proton,
neutron, and isoscalar nucleon) as well as the independently measured ratios of these
cross sections.
For the moment, our global likelihood analysis does not include the experimental data
for the  (quasi)elastic reactions with $\Delta Y\ne0$, the reactions of single-meson
production and NC induced (exclusive and inclusive) reactions. However, the bulk of the
data involved into the analysis is already rather representative and (more important)
more self-consistent in comparison with the data for the single- and multi-hadron
neutrinoproduction and the NC reactions of any kind. Hence we guess it is sufficient
for preliminary practical conclusions.

\section{Theoretical models}
  \label{Theoretical models}

\subsection{Quasielastic scattering}
     \label{Quasielastic scattering}

For the ${\nu}n\to\mu^-p$ and $\overline{\nu}p\to\mu^+n$ cross sections we use
the standard result (see, e.g., Ref.~\cite{LlewellynSmith:72}) neglecting the
second-class current contributions. 
For the elastic electromagnetic form factors $G_E^{p,n}$ and $G_M^{p,n}$
we apply the QCD Vector Meson model by Gari and Kr\"uempelmann~\cite{Gari:92}
extended and fine-tuned by Lomon~\cite{Lomon:02} to match the current and consistent
earlier experimental data derived using Rosenbluth separation and polarization transfer
techniques. More explicitly, we explore the so-called ``GKex(02S)'' version of the
model advocated by Lomon. 
At 4-momentum transfer $Q^2$ below $10-15~\text{GeV}^2$, the GKex(02S) model is
very close numerically to the PTD (polarization transfer data based) version of
the popular ``BBA-2003'' inverse-polynomial parametrization by Budd
\emph{et~al.}~\cite{BBA-2003} obtained through a global fit to the world data on
the Sachs form factors, including the results of several more recent measurements.
Although the up-to-date experiments (see, e.g., numerous reports in
Ref.~\cite{Structure_of_Baryons:05}) do not contradict to both models, we prefer
the GKex(02S) model since it meets the requirements of dispersion relations and QCD
asymptotics at low and high $Q^2$, while the BBA-2003 PTD fit has an unphysical behaviour
when extrapolated to high $Q^2$ (a typical drawback of polynomial approximations).

For the axial and pseudoscalar form factors we use the conventional
representations~\cite{LlewellynSmith:72}
\[
F_A\left(Q^2\right)=F_A(0)\left(1+\frac{Q^2}{M^2_A}\right)^{-2}
\]
and
\[
F_P\left(Q^2\right)=\frac{2M_N^2}{m^2_\pi+Q^2}F_A\left(Q^2\right),
\]
with $F_A(0)=g_A=-1.2695$~\cite{Eidelman:04}.
The currently available experimental data on the axial mass, $M_A=M_A^{\text{QES}}$,
show very wide spread, from roughly $0.6$ to $1.2~\text{GeV}/c^2$~\cite{Bernard:01}.
Today, it is the main source of uncertainties in the QES cross sections.
Since the pseudoscalar contribution enters into the cross sections multiplied
by $(m_{e,\mu,\tau}/M_N)^2$, it is substantial for neutrinoproduction of $\tau$ leptons
but small for electron and muon production; hence the related uncertainty is not
important for the present study.

Since the major part of the experimental data on QES obtained for heavy nuclear targets
was not corrected for nuclear effects, one have to take these into account in calculations.
We apply the simple Pauli factor since its effect for the total cross sections is not
essentially different from that evaluated with the more sophisticated approaches.

\subsection{Resonance single-pion production}
     \label{Resonance single-pion production}

In order to describe the single-pion neutrinoproduction through baryon resonances we
use an extended version of the model by Rein and Sehgal (RS)~\cite{Rein:81,Rein:87}.
The RS model, being one of the most circumstantial and approved phenomenological
approaches to calculating the RES cross sections, is now incorporated into essentially
all Monte Carlo neutrino event generators developed for both accelerator and
astroparticle experiments.
Our extension~\cite{Kuzmin:04,Kuzmin:05} takes into account the final lepton
mass~\cite{Footnote_Lepton_Spin} and is based upon a covariant form of the charged
leptonic current with definite lepton helicity.
In the present calculations, we use the same set of 18th interfering nucleon
resonances with masses below $2~\text{GeV}/c^2$ as in Ref.~\cite{Rein:81} but
with all relevant input parameters updated according to the current
data~\cite{Eidelman:04}.
Significant factors (normalization coefficients etc.) estimated in Ref.~\cite{Rein:81}
numerically are recalculated by using the new data and a more accurate integration
algorithm.

The relativistic quark model of Feynman, Kislinger, and Ravndal~\cite{Feynman:71}
adopted in the RS approach unambiguously determines the structure of the transition
amplitudes involved into the calculation and the only unknown structures are the
vector and axial-vector transition form factors $G^{V,A}\left(Q^2\right)$.
In Ref.~\cite{Rein:81} they are assumed to have the form 
\begin{equation}\label{G_VA}
G^{V,A}\left(Q^2\right)\propto\left(1+\frac{Q^2}{4M_N^2}\right)^{1/2-n}
\left(1+\frac{Q^2}{M_{V,A}^2}\right)^{-2}
\end{equation}
with the ``standard'' value of the vector mass $M_V=0.84~\text{GeV}/c^2$
(that is the same as in the naive dipole parametrization of the elastic vector
form factor)~\cite{Footnote_M_V}.
The axial mass $M_A=M_A^{\text{RES}}$ (which was fixed to be $0.95~\text{GeV}/c^2$
in the basic model) will be a free parameter in the present study.
The integer $n$ in the first (``ad hoc'') factor of Eq.~\eqref{G_VA}
is the number of oscillator quanta present in the final resonance.

To compensate for the difference between the $SU_6$ predicted value ($-5/3$)
and the experimental value for the nucleon axial-vector coupling $g_A$,
Rein and Sehgal introduced a renormalization factor $Z=0.75$.
In order to adjust the renormalization to the current world averaged value
$g_A=-1.2695\pm0.0029$~\cite{Eidelman:04} (assuming $g_V=1$) we have adopted
$Z=0.762$.

Another essential ingredient of the RS approach is the nonresonance background
(NRB) for which we use the ansatz suggested in Ref.~\cite{Rein:81}. The NRB
contribution is important for description of the existing data on the reactions
$         {\nu}_{\mu}n\to\mu^-n\pi^+$,
$         {\nu}_{\mu}n\to\mu^-p\pi^0$,
$\overline{\nu}_{\mu}p\to\mu^+p\pi^-$, and
$\overline{\nu}_{\mu}p\to\mu^+n\pi^0$.
Therefore it must be taken into account also in the RES contribution to the
total cross section if $W_{\text{cut}}^{\text{RES}} \le W_{\text{cut}}^{\text{DIS}}$.
It is not so obvious for the opposite (and by no means unphysical) case,
$W_{\text{cut}}^{\text{RES}}>W_{\text{cut}}^{\text{DIS}}$, since the DIS contribution
partially accounts for the NRB. So, it would be natural in this case to consider the
NRB as an additional ``free parameter'' of the likelihood analysis.
However, in this paper we pass over this complication and include the NRB
contribution into all variants of the fit.

\begin{figure}[htb]
\includegraphics[width=\linewidth]{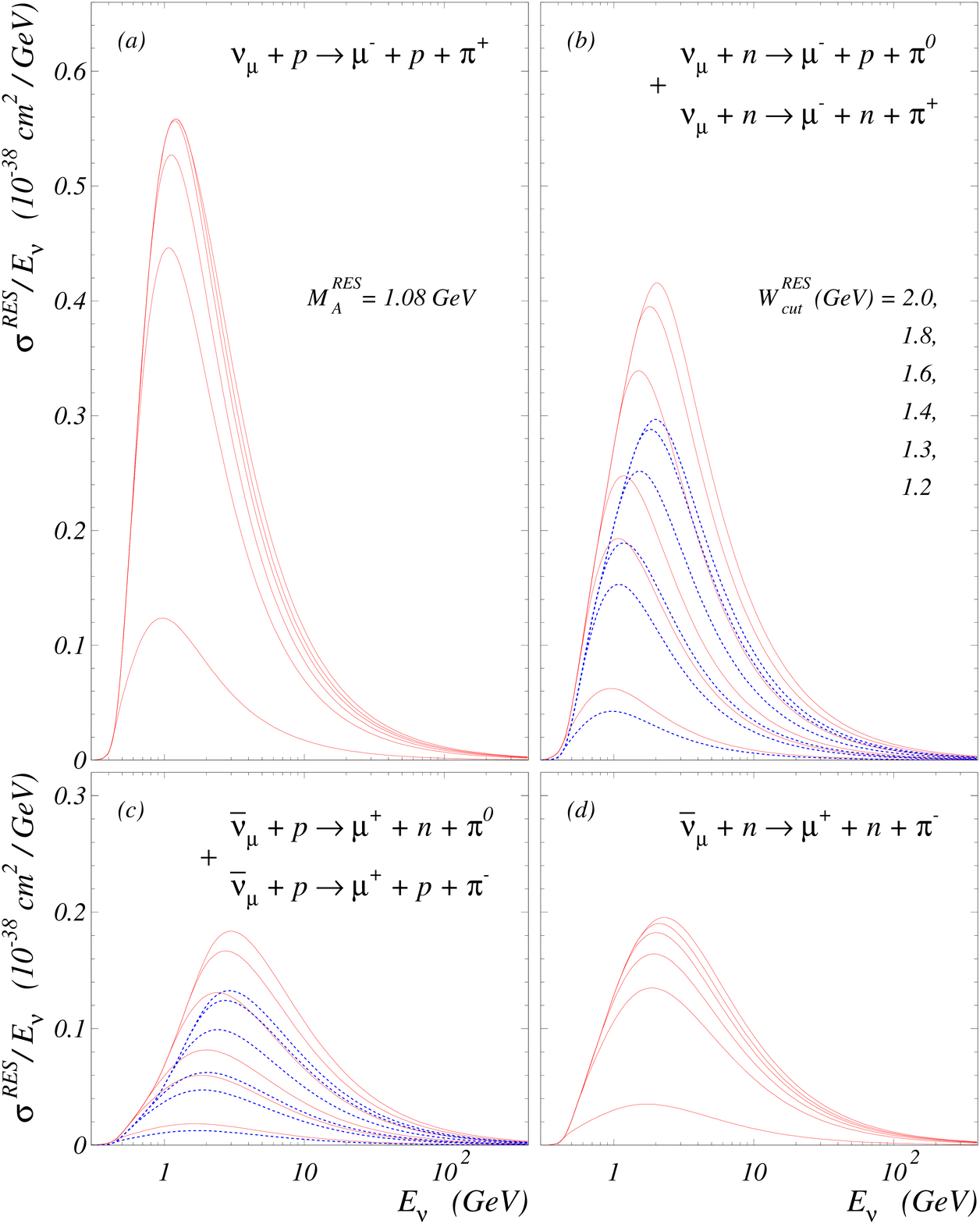}
\caption{The total CC single-pion production cross sections divided by
         neutrino energy evaluated with the extended RS model using
         $M_A^{\text{RES}}=1.08~\text{GeV}/c^2$ for the six values of
         the cutoff parameter
         $W_{\text{cut}}^{\text{RES}}=1.2, 1.3, 1.4, 1.6, 1.8, 2.0$~GeV
         (from bottom to top curves in each panel). The solid and dashed
         curves correspond to the cross sections calculated with and without
         the NRB contributions, respectively.
\label{Fig:SlopeRES}}
\end{figure}

Figure~\ref{Fig:SlopeRES} shows the RES contributions into the total CC
cross sections (divided by neutrino energy) evaluated by using the extended
RS model. In this example, we use $M_A^{\text{RES}}=1.08~\text{GeV}/c^2$,
the best fit value obtained from the recent analysis of the BNL 7-foot bubble
chamber deuterium experiment~\cite{Furuno:03} (hereafter referred to as ``BNL-2002'')
based on the total event sample of 1.8~M pictures (held two periods of runs in
1976-77 and 1979-80). The curves in panels (b) and (c) are for
the sums of the cross sections for the processes indicated in the legends.
The solid and dashed curves in these panels correspond to the calculation with
and without the NRB contributions, respectively. The calculations are done for
the six values of the cutoff parameter
$W_{\text{cut}}^{\text{RES}}=1.2, 1.3, 1.4, 1.6, 1.8, 2.0$~GeV;
clearly the cross sections decrease with decreasing the cutoff.

The next and more substantial drawback of the present study is in neglecting the
nuclear corrections for the RES (as well as for the DIS) contribution.
A justification is in the fact that these effects were subtracted in a certain part
of the total cross section data while the necessary information is unavailable for
another part of the data. We intend to remove this drawback in future study.

\subsection{Deep inelastic scattering}
     \label{Deep inelastic scattering}

The DIS CC ${\nu}_{\mu}N$ and $\overline{\nu}_{\mu}N$ differential cross sections
are represented by the standard set of five structure functions $F_i=F_i(x,Q^2)$
(see, e.g., Refs.~\cite{Kretzer:02,Kretzer:04a}):
\begin{equation}\label{KR_sigma}
\frac{d^2\sigma^{\text{DIS}}_{\nu(\bar{\nu})}}{dxdy}
=\frac{G_F^2M_NE_\nu}{\pi(1+Q^2/M_W^2)^2}
\sum_{i=1}^5A_i\left(x,y,E_\nu\right)F_i\left(x,Q^2\right),
\end{equation} 
where $x$ and $y$ are the usual DIS kinematic variables.
The coefficient functions $A_i$ are
\begin{equation}\label{A_i}
\begin{aligned}
A_1&=y\left(xy+\frac{m_{\mu}^2}{2M_NE_{\nu}}\right),                              \\
A_2&=1-\left(1+\frac{M_Nx}{2E_{\nu}}\right)y-\frac{m_{\mu}^2}{4E_{\nu}^2},        \\
A_3&=\pm y\left[x\left(1-\frac{y}{2}\right)-\frac{m_{\mu}^2}{4M_NE_{\nu}}\right], \\
A_4&=\frac{m_{\mu}^2}{2M_NE_{\nu}}\left(y+\frac{m_{\mu}^2}{2M_NE_{\nu}x}\right),  \\
A_5&=-\frac{m_{\mu}^2}{M_NE_{\nu}}.
\end{aligned}
\end{equation} 
The functions $F_1$ and $F_2$ are related through the measurable structure function
$R=F_L/(2xF_1)=\sigma_L/\sigma_T$, the ratio of longitudinal and transverse cross
sections in DIS:
\begin{equation}\label{ExactCallan-GrossRelation}
\mathfrak{D}F_2=2xF_1,
\quad
\mathfrak{D}=\frac{1}{1+R}\left(1+\frac{Q^2}{{\nu}^2}\right),
\end{equation}
where $\nu=yE_\nu$. In order to satisfy Eq.~\eqref{ExactCallan-GrossRelation} and,
simultaneously, the collinear parton model (PM) limit that is the Callan-Gross relation
($F_2^{\text{PM}}\to2xF_1^{\text{PM}}$ as $Q^2\to\infty$ or $\mathfrak{D}\to1$),
the exact structure functions $F_{1,2}$ must be related to those in the PM limit,
$F_{1,2}^{\text{PM}}$, as
\begin{equation}
F_1=\left(1-a+a\mathfrak{D}\right)F_1^{\text{PM}},
\quad
F_2=\left[a+(1-a)/\mathfrak{D}\right]F_2^{\text{PM}}.
\end{equation}
Till the function $a=a(x,Q^2)$ is not specified, these relations are the most
general.
There are two simplest limiting possibilities for $a$:
$a=0$ ($F_1=F_1^{\text{PM}}$, $F_2=F_2^{\text{PM}}/\mathfrak{D}$) and
$a=1$ ($F_1=\mathfrak{D}F_1^{\text{PM}}$, $F_2=F_2^{\text{PM}}$).
Our analysis of the experimental data described in the next section and testing
many models for the parton density functions (PDF) suggests that the ``$a=1$ model''
works quite satisfactory. Hereafter we will discuss just this particular case.

For the structure function $R$ we use a combination of two up-to-date
parametrizations: inside the nucleon resonance region $1.15<W^2<3.9~\text{GeV}^2$
and $0.3<Q^2<5.0~\text{GeV}^2$ we apply the recent precision result of the Jefferson
Lab Hall C E94-110 Collaboration~\cite{Liang:04,Liang:05}; outside this region we
apply the $R_a$ version of the accurate 6-parameter fit to the world data on $R$
proposed by the SLAC E-143 Collaboration~\cite{Abe:98}.
The two parametrizations are sewn by a 2D B-spline in the boundary of the kinematic
regions.

\begin{figure}[htb]
\includegraphics[width=\linewidth]{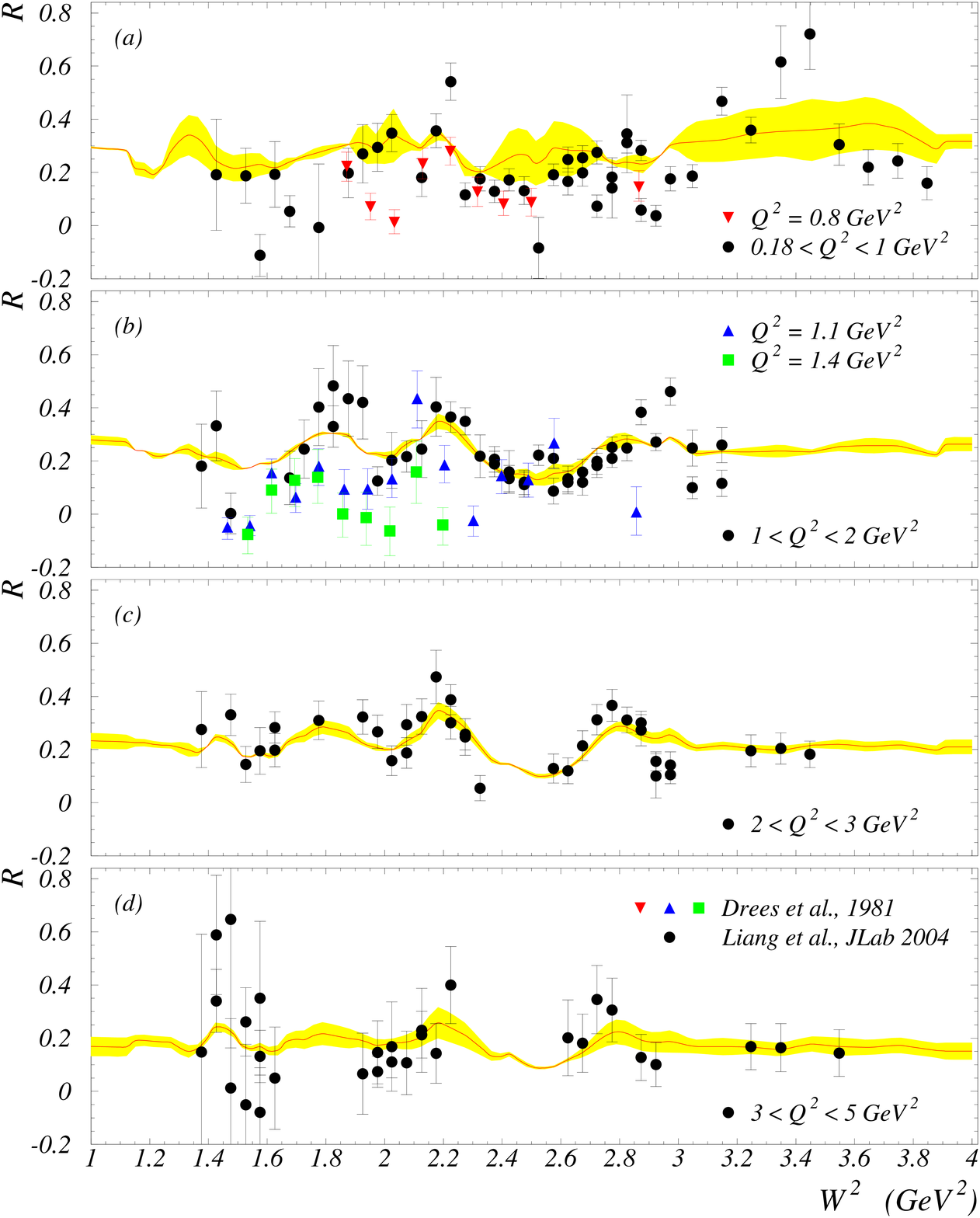}
\caption{The structure function $R=\sigma_L/\sigma_T$ vs.\ $W^2$ obtained by
         the Rosenbluth analysis of the inclusive $ep$ cross sections measured
         at the JLab Hall C experiment~\cite{Liang:04} for the $Q^2$ ranges
         indicated in the panels.
         Also shown are the results of several earlier experiments on
         $ep$ scattering in the resonance region converted to $R$ in
         Ref.~\cite{Dress:81} for $Q^2=0.8$ (a), $1.1$ and $1.4~\text{GeV}^2$ (b).
         The bands are evaluated by using the model for $R$ described in the text
         and by varying $Q^2$ within the corresponding ranges; the curves are
         for the $R$ averaged over these ranges.
\label{Fig:RW2_Liang+R_a_spline}}
\end{figure}

In Fig.~\ref{Fig:RW2_Liang+R_a_spline} we show a comparison of the described model
with the data from JLab~\cite{Liang:04} and the results of the $\sigma_L$ and
$\sigma_T$ separation performed by Dress \emph{et~al.}~\cite{Dress:81} and based
on the data of many older measurements of the $ep$ cross sections in the resonance
region.
The filled bands in the figure are obtained by varying $Q^2$ within the ranges
$0.18-1~\text{GeV}^2$ (a), $1-2~\text{GeV}^2$ (b),
$2-3~\text{GeV}^2$ (c), and $3-5~\text{GeV}^2$ (d).

Since the JLab fit has been obtained from the data on $ep$ scattering,
we corrected it to the ${\nu}N$ scattering and tested by using the QCD based
Altarelli-Martinelli equation~\cite{Altarelli:78}. In fact, the difference between
$R^{(e,\mu)}$ and $R^{(\nu,\overline{\nu})}$ is practically negligible
within the relevant kinematic region below the charm production threshold and
small above the threshold. 

From Eqs.~\eqref{A_i} and the exact ${\nu}N$ kinematics it follows that
\[
A_4<\frac{m_{\mu}^2}{2M_NE_\nu}\left(1-\frac{m_{\mu}}{E_\nu}\right)
<\frac{m_{\mu}}{2M_N}
\quad\text{and}\quad
|A_5|<\frac{m_{\mu}}{M_N}.
\]
Due to this suppression and in view of the scale of the functions $F_4$ and $F_5$
followed from the NLO pQCD plus target mass calculations~\cite{Kretzer:02}, the
$A_4F_4$ and $A_5F_5$ terms in Eq.~\eqref{KR_sigma} can only be significant
near the reaction threshold~\cite{Footnote_Tau}.
Hence the structure functions $F_{4,5}$ can be estimated roughly, by using the
approximate relations valid in the PM limit with massless quarks:
\[
F_4\approx\frac{1}{2}\left(\frac{F_2}{2x}-F_1\right)=
\frac{1}{2}\left(\frac{1}{\mathfrak{D}}-1\right)F_1
\]
and
\[
F_5\approx\frac{F_2}{2x}=\frac{F_1}{\mathfrak{D}}.
\]

\begin{figure}[htb]
\includegraphics[width=\linewidth]{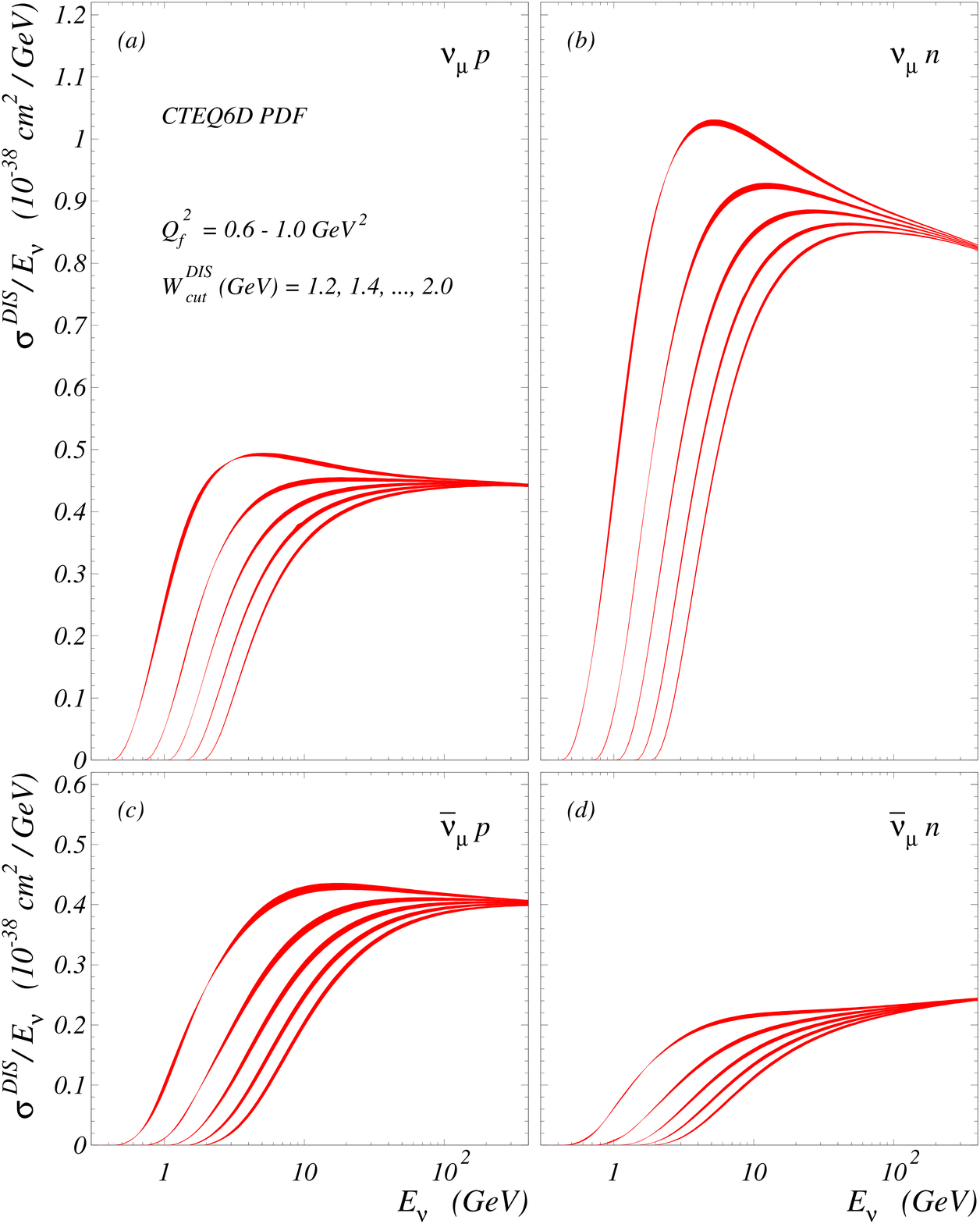}
\caption{The total DIS $\nu_{\mu}p$, $\nu_{\mu}n$, $\overline{\nu}_{\mu}p$,
         and $\overline{\nu}_{\mu}n$ CC cross sections divided by neutrino energy
         evaluated with the CTEQ~6D NLO PDFs for the five values of the
         cutoff parameter $W_{\text{cut}}^{\text{DIS}}=1.2$ to 2.0~GeV
         from top to bottom with steps of 0.2~GeV. The widths of the bands
         correspond to variation of the freezing parameter $Q_f^2$ from $0.6$
         to $1.0~\text{GeV}^2$.
\label{Fig:SlopeDIS-6.02}}
\end{figure}

The PDF contributions into all structure functions are divided, in the standard
fashion, onto ``non charm production'' (ncp) and ``charm production'' (cp) parts:
\[
q^{\text{ncp}}=q^{\text{ncp}}(x_N,Q^2)
\quad\text{and}\quad
q^{\text{cp}}=q^{\text{cp}}(\xi,Q^2),
\]
where
$x_N=2x/\left(1+\sqrt{1+Q^2/\nu^2}\right)$ is the Nachtmann variable,
$\xi=x_N\left(1+{m_c^2}/{Q^2}\right)$ is the collinear limit of the light-cone
variable with massless $u$, $d$, and $s$ quarks, and $m_c=1.3~\text{GeV}/c^2$
is the mass of $c$ quark. The $b$ and $t$ quark contributions are neglected.

We have tested several popular PDF models but in this paper we only discuss the
results obtained with the latest version of CTEQ~6D NLO PDF set with four
flavors (standard DIS scheme, version~6.12, December 14, 2004)~\cite{CTEQ6}.

Figure~\ref{Fig:SlopeDIS-6.02} shows the total DIS $\nu_{\mu}p$, $\nu_{\mu}n$,
$\overline{\nu}_{\mu}p$, and $\overline{\nu}_{\mu}n$ CC cross sections divided
by neutrino energy evaluated with the CTEQ~6D NLO PDFs for the five values of the
cutoff parameter $W_{\text{cut}}^{\text{DIS}}=1.2, 1.4, 1.6, 1.8$, and 2.0~GeV;
Clearly, the cross sections increase with decreasing of the cutoff.
Since the CTEQ~6D PDFs cannot be extrapolated to the exact kinematic boundaries,
we have to freeze $Q^2$ below some value $Q_f^2$. In Fig.~\ref{Fig:SlopeDIS-6.02},
this value varies within the range $0.6$ to $1.0~\text{GeV}^2$ and the widths
of the bands reflect the corresponding variations of the DIS cross sections.
The $Q_f$ dependence is in general nonmonotonic and diminishes with increasing
the cutoff value.
In the present likelihood analysis, the $Q^2$ variable is freezing below
$Q_f^2=0.8~\text{GeV}^2$.
The error introduced by this approximation is estimated to be less than 1-2\%
that is small in comparison with the uncertainties of the experimental data and
indetermination in other phenomenological parameters.

\section{Data set}
  \label{Data set}

We have examined and classified all available experimental data on the QES and total
CC ${\nu}N$ and $\overline{\nu}N$ cross sections as well as independently measured
relative quantities like the ratios
$\sigma_{{\nu}n}/\sigma_{{\nu}p}$,
$\sigma_{\overline{\nu}n}/\sigma_{\overline{\nu}p}$,
$\sigma_{\overline{\nu}p}/\sigma_{\nu p}$, and so on.
Published results from the relevant experiments at
 ANL~\cite{Kustom:69,Mann:73,Barish:77,Barish:79},
 BNL~\cite{Fanourakis:80,Baltay:80,Baker:81,Jacques:81,Baker:82},
FNAL~\cite{Barish:68,Benvenuti:73,Benvenuti:74,Imlay:74,Barish:75PRL,Benvenuti:76,%
           Sciulli:78,Barish:77PRL,Barish:78,Efremenko:79,Hanlon:80,Barish:81,%
           Kitagaki:82,Kitagaki:83,Baker:83PRL,Taylor:83,Asratyan:84a,Asratyan:84b,%
           MacFarlane:84,Auchincloss:90,Seligman:97,Suwonjandee:04,Tzanov:05}
LANL~\cite{Auerbach:02},
CERN~\cite{Young:67,Budagov:69LNC,Budagov:69PLB,Myatt:71,Eichten:73PLBa,%
          Eichten:73PLBb,Bonetti:77,Holder:77,Blietschau:78,Lerche:78,Musset:78,%
          Armenise:79,Pohl:79,Colley:79,Erriquez:79,Ciampolillo:79,Armenise:81,%
          Jonker:81,Morfin:81,Bosetti:82,Abramowicz:83,Abramowicz:84,Parker:84,%
          Allasia:84,Shotton:85,Aderholz:86,Allaby:86,Berge:87,Allaby:88,%
          Allasia:90,Kayis-Topaksu:02},
and
IHEP~\cite{Asratyan:78,Baranov:79,Vovenko:79,Makeev:81,Belikov:82,Belikov:85,%
           Grabosch:88,Brunner:89ZPCa,Brunner:89ZPCb,Ammosov:92,Anikeev:96}
are included dating from the end of sixties to the present day, covering $\nu_{\mu}$,
$\overline{\nu}_{\mu}$, $\nu_e$, and $\overline{\nu}_e$ beams on a variety of hydrogen
and nuclear targets, with energies from the thresholds to about 350~GeV. A detailed
description of our database will be published elsewhere.
Here we briefly depict the most important points.

Not all the collected data are involved into the analysis.
We excluded from the fit:
\begin{itemize}
\item
the experimental results which are undoubtedly obsolete, superseded or reconsidered
(due to increased statistics, revised normalization, etc.) in the posterior reports
of the same Collaborations;
\item
the datasets which are a transformation of the others derived from the same experimental
samples (for instance, we used either the cross section $\sigma$ or the ``slope''
$\sigma/E_\nu$ measured in the same experiment);
\item
the cross sections, slopes, and ratios averaged over a wide energy range when
the energy-binned dataset is available.
\end{itemize}
We quenched a wish to reject the results seeming self-contradictory or being in
obvious disagreement with the major dataset. A few exceptions and particular
cases will be expounded in the next section. 

If only the bounds of the energy bin were available, we either averaged
the data (and the relevant calculated quantity) over the bin or estimated the
mean energy from the (anti)neutrino beam spectrum (when the necessary information
was accessible from the original paper or another description of the experiment).
The statistical and systematic errors of the data were always summed up quadratically.

\section{Likelihood analysis}
  \label{Likelihood analysis}

The four above-mentioned parameters $M_A^{\text{QES}}$, $M_A^{\text{RES}}$,
$W_{\text{cut}}^{\text{RES}}$, and $W_{\text{cut}}^{\text{DIS}}$ involved into
the merging of the QES, RES, and DIS contributions (Sect.~\ref{Theoretical models})
were fitted to the described data by using the CERN function minimization and error
analysis package ``MINUIT'' (version 94.1)~\cite{James:94}.
In order to test validity of the dataset and the fitting procedure, we have examined
many variants of 1-, 2-, 3-, and 4-parameter fits taking care of getting the correct
correlation coefficients printed by MINUIT.
Illustrative results of this analysis are listed in Tables~%
\ref{Tab:FIT-16.02-f-6.02-2-0.80_1}, \ref{Tab:FIT-16.02-f-6.02-2-0.80_2} and
\ref{Tab:FIT-16.02-f-6.02-2-0.80_3} together with the obtained values of
$\chi^2$ per number of degrees of freedom (NDF). The number of significant
digits shown in the last columns of Tables~\ref{Tab:FIT-16.02-f-6.02-2-0.80_2}
and \ref{Tab:FIT-16.02-f-6.02-2-0.80_3} is more than needed; we only keep these
to clarify the $\chi^2$ minima.

The first column in each table is for designation of different exercises of the fit.
The numbers in bold-face correspond to the fixed trial values of the parameters
used as inputs. The errors of the output parameters correspond to the usual
one-standard-deviation ($1\sigma$) errors (MINUIT default)~\cite{Footnote_Big_Errors}.
\begin{table}[htb]
\caption{One-parameter fits with the corresponding $\chi^2$
         per $\text{NDF}=669$. Trial parameters are bold-faced.}
\bigskip
\begin{tabular}{c|c|c|c|c|c}
\toprule 
Fit & $M^{\text{QES}}_A$             &$M^{\text{RES}}_A$                 &
 $W^{\text{RES}}_{\text{cut}}$       &$W^{\text{DIS}}_{\text{cut}}$      &
                                                      $\dfrac{\chi^2}{\text{NDF}}$ \\
    & (GeV)                          & (GeV)                             & (GeV)
    & (GeV)                          &                                             \\
\colrule 
  &\u{0.8}      &     \u{1.08}       &\multicolumn{2}{c|}{$1.47\pm0.02$} & $1.67$  \\
  &\u{0.9}      &     \u{1.08}       &\multicolumn{2}{c|}{$1.50\pm0.01$} & $1.52$  \\
A1&\u{1.0}      &     \u{1.08}       &\multicolumn{2}{c|}{$1.53\pm0.02$} & $1.47$  \\
  &\u{1.1}      &     \u{1.08}       &\multicolumn{2}{c|}{$1.56\pm0.01$} & $1.56$  \\
  &\u{1.2}      &     \u{1.08}       &\multicolumn{2}{c|}{$1.58\pm0.02$} & $1.80$  \\
\colrule 
  &\multicolumn{2}{c|}{\u{0.8}}      &\u{2.0}      &     $1.47\pm0.01$   & $1.68$  \\
  &\multicolumn{2}{c|}{\u{0.9}}      &\u{2.0}      &     $1.52\pm0.01$   & $1.52$  \\
B1&\multicolumn{2}{c|}{\u{1.0}}      &\u{2.0}      &     $1.58\pm0.01$   & $1.47$  \\
  &\multicolumn{2}{c|}{\u{1.1}}      &\u{2.0}      &     $1.64\pm0.01$   & $1.56$  \\
  &\multicolumn{2}{c|}{\u{1.2}}      &\u{2.0}      &     $1.71\pm0.01$   & $1.82$  \\
\botrule 
\end{tabular}
\label{Tab:FIT-16.02-f-6.02-2-0.80_1}
\end{table}

\begin{table}[htb]
\caption{Two-parameter fits with the corresponding $\chi^2$
         per $\text{NDF}=668$. Trial parameters are bold-faced.}
\bigskip
\begin{tabular}{c|c|c|c|c|c}
\toprule 
Fit & $M^{\text{QES}}_A$             &$M^{\text{RES}}_A$                 &
 $W^{\text{RES}}_{\text{cut}}$       &$W^{\text{DIS}}_{\text{cut}}$      &
                                                      $\dfrac{\chi^2}{\text{NDF}}$ \\
    & (GeV)                          & (GeV)                             & (GeV)
    & (GeV)                          &                                             \\
\colrule 
A2&$0.99\pm0.02$&     \u{1.08}       &\multicolumn{2}{c|}{$1.53\pm0.02$} & $1.469$ \\
\colrule 
B2&\multicolumn{2}{c|}{$0.96\pm0.01$}&$1.31\pm0.01$&     \u{1.4}         & $1.484$ \\
\colrule 
  &\u{0.8}      &$1.11\pm0.08$       &$1.28\pm0.01$&     \u{1.4}         & $1.631$ \\
  &\u{0.9}      &$1.10\pm0.08$       &$1.28\pm0.01$&     \u{1.4}         & $1.499$ \\
C2&\u{1.0}      &$1.09\pm0.07$       &$1.28\pm0.01$&     \u{1.4}         & $1.493$ \\
  &\u{1.1}      &$1.08\pm0.07$       &$1.28\pm0.01$&     \u{1.4}         & $1.635$ \\
  &\u{1.2}      &$1.07\pm0.06$       &$1.29\pm0.01$&     \u{1.4}         & $1.943$ \\
\colrule 
  &\u{0.8}      &$0.91\pm0.04$       &\multicolumn{2}{c|}{$1.41\pm0.02$} & $1.649$ \\
  &\u{0.9}      &$0.97\pm0.04$       &\multicolumn{2}{c|}{$1.46\pm0.02$} & $1.507$ \\
D2&\u{1.0}      &$1.03\pm0.04$       &\multicolumn{2}{c|}{$1.51\pm0.02$} & $1.468$ \\
  &\u{1.1}      &$1.07\pm0.04$       &\multicolumn{2}{c|}{$1.55\pm0.02$} & $1.558$ \\
  &\u{1.2}      &$1.10\pm0.04$       &\multicolumn{2}{c|}{$1.59\pm0.02$} & $1.801$ \\
\colrule 
  &\u{0.8}      &$1.00\pm0.04$       &\u{2.0}      &$1.53\pm0.02$        & $1.651$ \\
  &\u{0.9}      &$1.04\pm0.04$       &\u{2.0}      &$1.57\pm0.02$        & $1.505$ \\
E2&\u{1.0}      &$1.08\pm0.04$       &\u{2.0}      &$1.61\pm0.02$        & $1.469$ \\
  &\u{1.1}      &$1.11\pm0.04$       &\u{2.0}      &$1.65\pm0.02$        & $1.566$ \\
  &\u{1.2}      &$1.14\pm0.04$       &\u{2.0}      &$1.68\pm0.02$        & $1.814$ \\
\colrule 
  &\u{0.8}      &     \u{1.08}       &$1.28\pm0.01$&$1.38\pm0.01$        & $1.629$ \\
  &\u{0.9}      &     \u{1.08}       &$1.77\pm0.09$&$1.56\pm0.02$        & $1.500$ \\
F2&\u{1.0}      &     \u{1.08}       &$1.70\pm0.14$&$1.57\pm0.03$        & $1.465$ \\
  &\u{1.1}      &     \u{1.08}       &$1.34\pm0.03$&$1.49\pm0.01$        & $1.551$ \\
  &\u{1.2}      &     \u{1.08}       &$1.37\pm0.04$&$1.53\pm0.02$        & $1.787$ \\
\botrule 
\end{tabular}
\label{Tab:FIT-16.02-f-6.02-2-0.80_2}
\end{table}

\begin{table}[htb]
\caption{Three- and four-parameter fits with the corresponding $\chi^2$
         per NDF ($=667$ and $666$, respectively).
         Trial parameters are bold-faced.}
\bigskip
\begin{tabular}{c|c|c|c|c|c}
\toprule 
Fit & $M^{\text{QES}}_A$             &$M^{\text{RES}}_A$                 &
 $W^{\text{RES}}_{\text{cut}}$       &$W^{\text{DIS}}_{\text{cut}}$      &
                                                      $\dfrac{\chi^2}{\text{NDF}}$ \\
    & (GeV)                          & (GeV)                             & (GeV)
    & (GeV)                          &                                             \\
\colrule 
  &$0.93\pm0.01$&$1.06\pm0.07$       &$1.27\pm0.01$&      \u{1.35}       & $1.542$ \\
  &$0.95\pm0.01$&$1.09\pm0.08$       &$1.28\pm0.01$&      \u{1.40}       & $1.481$ \\
A3&$0.98\pm0.01$&$1.13\pm0.08$       &$1.30\pm0.02$&      \u{1.45}       & $1.461$ \\
  &$0.98\pm0.02$&$1.00\pm0.03$       &$1.55\pm0.04$&      \u{1.50}       & $1.467$ \\
  &$0.98\pm0.02$&$1.05\pm0.03$       &$1.69\pm0.07$&      \u{1.55}       & $1.463$ \\
\colrule 
B3&$0.98\pm0.02$&$1.02\pm0.04$       &\multicolumn{2}{c|}{$1.50\pm0.02$} & $1.468$ \\
\colrule 
  &$0.98\pm0.01$&$1.01\pm0.03$       &\u{1.50}     &$1.50\pm0.00$        & $1.468$ \\
  &$0.98\pm0.02$&$1.02\pm0.04$       &\u{1.55}     &$1.51\pm0.02$        & $1.466$ \\
  &$0.98\pm0.02$&$1.03\pm0.04$       &\u{1.60}     &$1.53\pm0.02$        & $1.464$ \\
  &$0.98\pm0.02$&$1.04\pm0.04$       &\u{1.65}     &$1.54\pm0.02$        & $1.463$ \\
  &$0.98\pm0.02$&$1.05\pm0.04$       &\u{1.70}     &$1.55\pm0.02$        & $1.463$ \\
C3&$0.98\pm0.02$&$1.06\pm0.04$       &\u{1.75}     &$1.56\pm0.02$        & $1.464$ \\
  &$0.98\pm0.02$&$1.06\pm0.04$       &\u{1.80}     &$1.57\pm0.02$        & $1.464$ \\
  &$0.98\pm0.02$&$1.06\pm0.04$       &\u{1.85}     &$1.58\pm0.02$        & $1.465$ \\
  &$0.98\pm0.02$&$1.07\pm0.04$       &\u{1.90}     &$1.59\pm0.02$        & $1.467$ \\
  &$0.98\pm0.02$&$1.07\pm0.04$       &\u{1.95}     &$1.59\pm0.02$        & $1.468$ \\
  &$0.98\pm0.02$&$1.07\pm0.04$       &\u{2.00}     &$1.60\pm0.02$        & $1.469$ \\
\colrule 
D3&$0.98\pm0.01$&     \u{1.08}       &$1.73\pm0.26$&$1.57\pm0.05$        & $1.464$ \\
\colrule 
  &\u{0.8}      &$1.10\pm0.08$       &$1.28\pm0.01$&$1.38\pm0.01$        & $1.631$ \\
  &\u{0.9}      &$1.11\pm0.08$       &$1.29\pm0.02$&$1.42\pm0.01$        & $1.497$ \\
E3&\u{1.0}      &$1.13\pm0.08$       &$1.30\pm0.02$&$1.46\pm0.01$        & $1.463$ \\
  &\u{1.1}      &$1.17\pm0.06$       &$1.31\pm0.02$&$1.50\pm0.01$        & $1.551$ \\
  &\u{1.2}      &$1.19\pm0.06$       &$1.32\pm0.02$&$1.53\pm0.01$        & $1.783$ \\
\colrule 
A4&$0.98\pm0.02$&$1.13\pm0.08$       &$1.30\pm0.02$&$1.45\pm0.01$        & $1.463$ \\
\botrule 
\end{tabular}
\label{Tab:FIT-16.02-f-6.02-2-0.80_3}
\end{table}

\begin{figure}[htb]
\includegraphics[width=\linewidth]{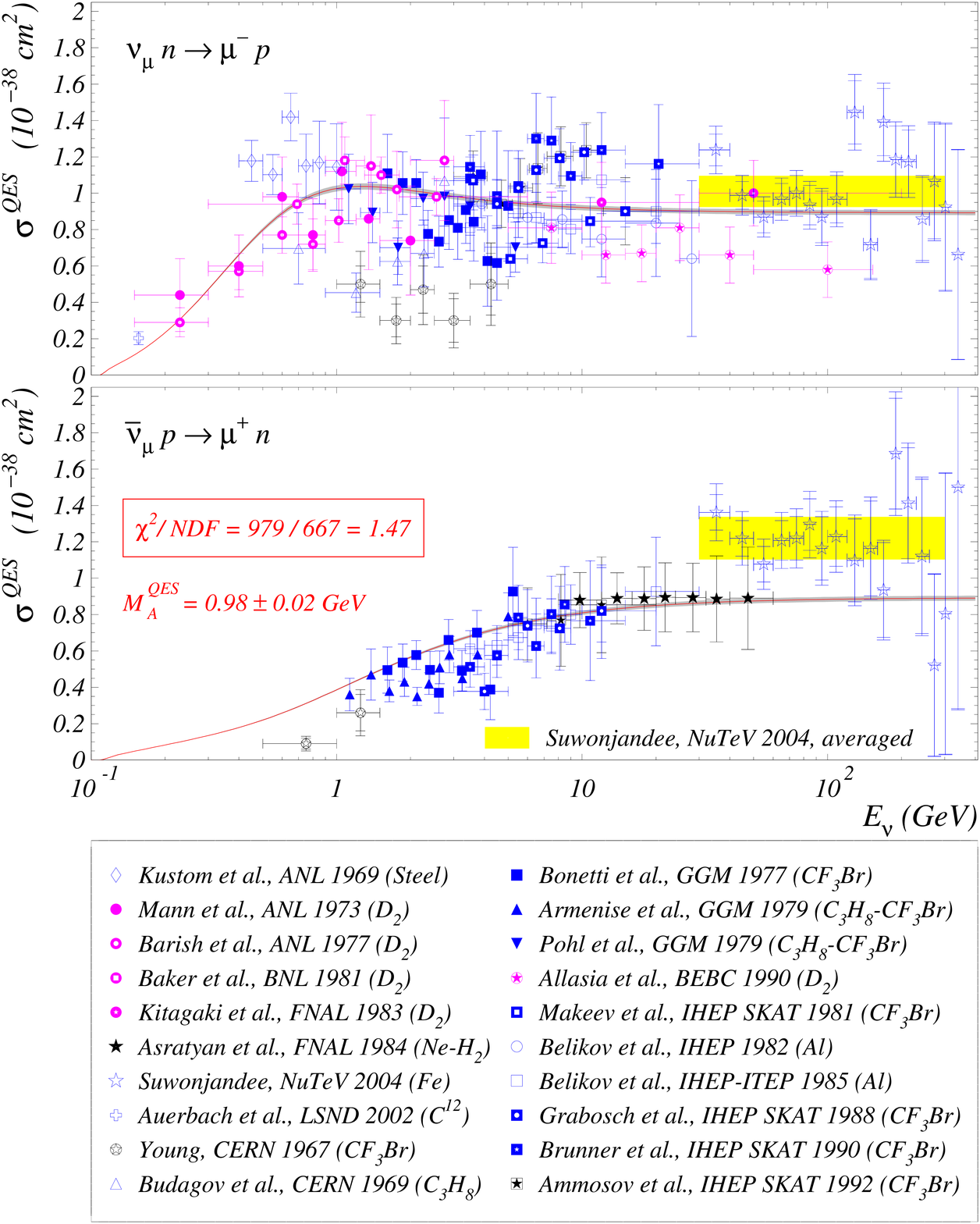}
\caption{Total QES cross sections measured by the experiments
         ANL~1969~\cite{Kustom:69},
         ANL~1973~\cite{Mann:73},
         ANL~1977~\cite{Barish:77},
         BNL~1981~\cite{Baker:81},
         FNAL~1983~\cite{Kitagaki:83},
         FNAL~1984~\cite{Asratyan:84b},
         NuTeV~2004~\cite{Suwonjandee:04},
         LSND~2002~\cite{Auerbach:02},
         CERN~1967~\cite{Young:67},
         CERN~1969~\cite{Budagov:69LNC},
         GGM~(Gargamelle)~1977~\cite{Bonetti:77},
         GGM~1979~\cite{Armenise:79,Pohl:79},
         BEBC~1990~\cite{Allasia:90},
         IHEP~SKAT~1981~\cite{Makeev:81},
         IHEP~1982~\cite{Belikov:82},
         IHEP-ITEP~1985~\cite{Belikov:85},
         IHEP~SKAT~1988~\cite{Grabosch:88},
         IHEP~SKAT~1990~\cite{Brunner:89ZPCb},
         and
         IHEP~SKAT~1992~\cite{Ammosov:92}.
         Both statistical and total errors are shown for the earliest
         low-energy data of CERN~\cite{Young:67} (excluded from the fit)
         and for the most current high-energy data of NuTeV~\cite{Suwonjandee:04}.
         The filled rectangles are for the NuTeV data (with the total error)
         averaged over the wide energy range 30 to 300 GeV.
         The data for nuclear targets (indicated in the parentheses in the legend)
         are converted to a free nucleon.
         The curves are for the QES cross sections calculated with the value
         of $M_A^{\text{QES}}=0.98~\text{GeV}/c^2$ obtained from the global B3 fit
         (see text).
         The narrow grey bands show the standard $1\sigma$ deviation from the
         best-fit curves.
\label{Fig:SigmaQES-Fit-16.02-f-6.02-2-0.80}}
\end{figure}

Visualization of the results is shown in Figs.~%
\ref{Fig:SigmaQES-Fit-16.02-f-6.02-2-0.80}--%
\ref{Fig:SlopeRatioSUM-npI-fit-16.02-f-6.02-2-0.80}
for the B3 variant of the fit which is preferable in our opinion.

\begin{figure}[htb]
\includegraphics[width=\linewidth]{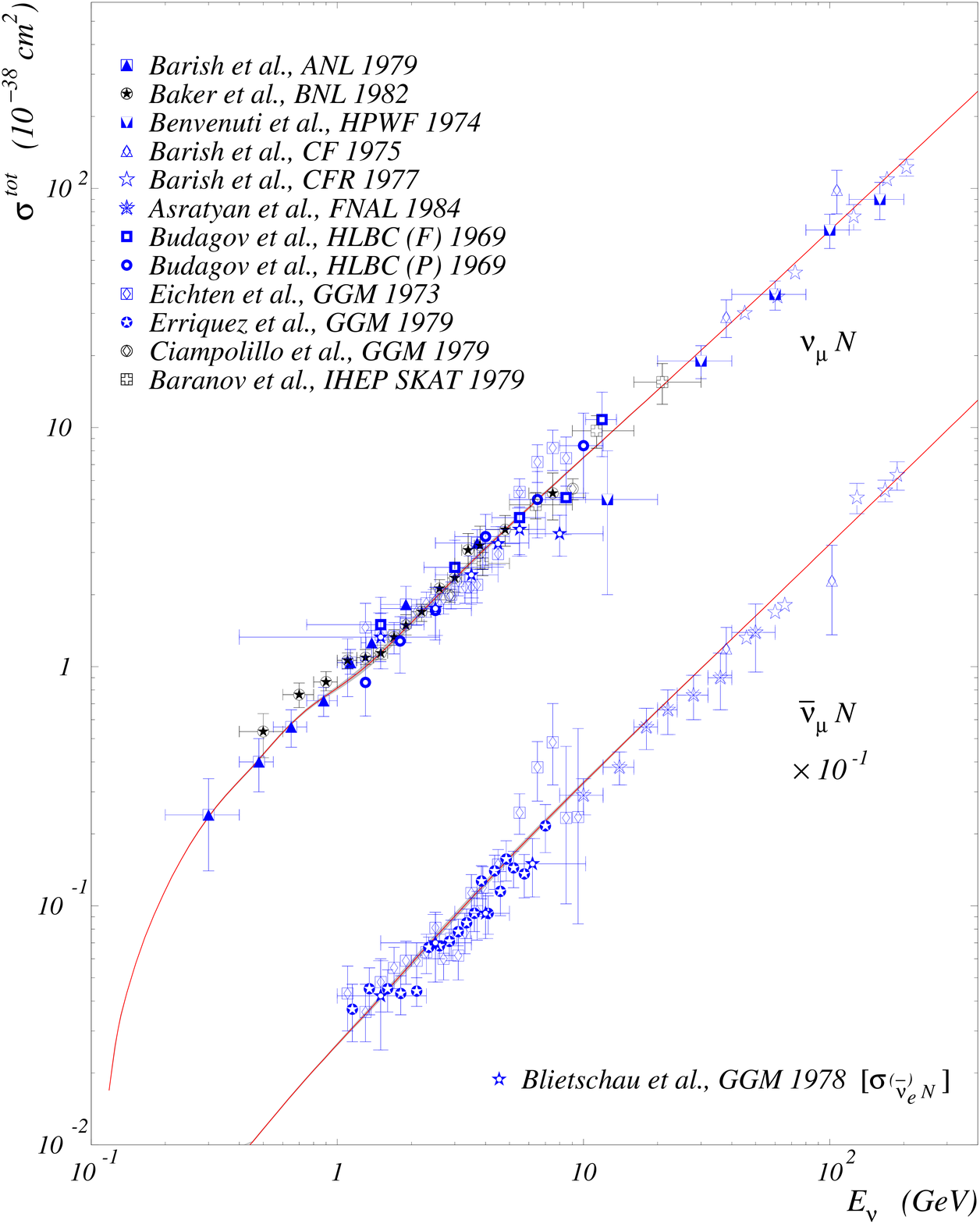}
\caption{Total CC cross sections for $\nu_\mu$ and $\overline{\nu}_\mu$
         scattering off an isoscalar nucleon measured by the experiments
         ANL~1979~\cite{Barish:79},
         BNL~1982~\cite{Baker:82},
         HPWF~1974~\cite{Benvenuti:74},
         CF~1975~\cite{Barish:75PRL},
         CFR~1977~\cite{Barish:77PRL},
         FNAL~1984~\cite{Asratyan:84b},
         HLBC~1969 (freon, 1963-64 and propane, 1967 runs)~\cite{Budagov:69PLB},
         GGM~1973~\cite{Eichten:73PLBa},
         GGM~1979~\cite{Erriquez:79,Ciampolillo:79},
         and
         IHEP~SKAT~1979~\cite{Baranov:79}.
         Also shown are the $\nu_eN$ and $\overline{\nu}_eN$ cross
         sections measured by the GGM~1978 experiment~\cite{Blietschau:78}.
         The antineutrino data are scaled with a factor of 0.1 for
         better visualization.
         The curves and bands show the cross sections calculated with
         the best-fitted values of $M_A^{\text{QES}}$, $M_A^{\text{RES}}$,
         and $W_{\text{cut}}^{\text{RES}}=W_{\text{cut}}^{\text{DIS}}$
         (see text and legend in
         Fig.~\protect\ref{Fig:SlopeSUM-I-FIT-16.02-f-6.02-2-0.80}).
\label{Fig:SigmaSUM-I-FIT-16.02-f-6.02-2-0.80}}
\end{figure}
\begin{figure}[htb]
\includegraphics[width=\linewidth]{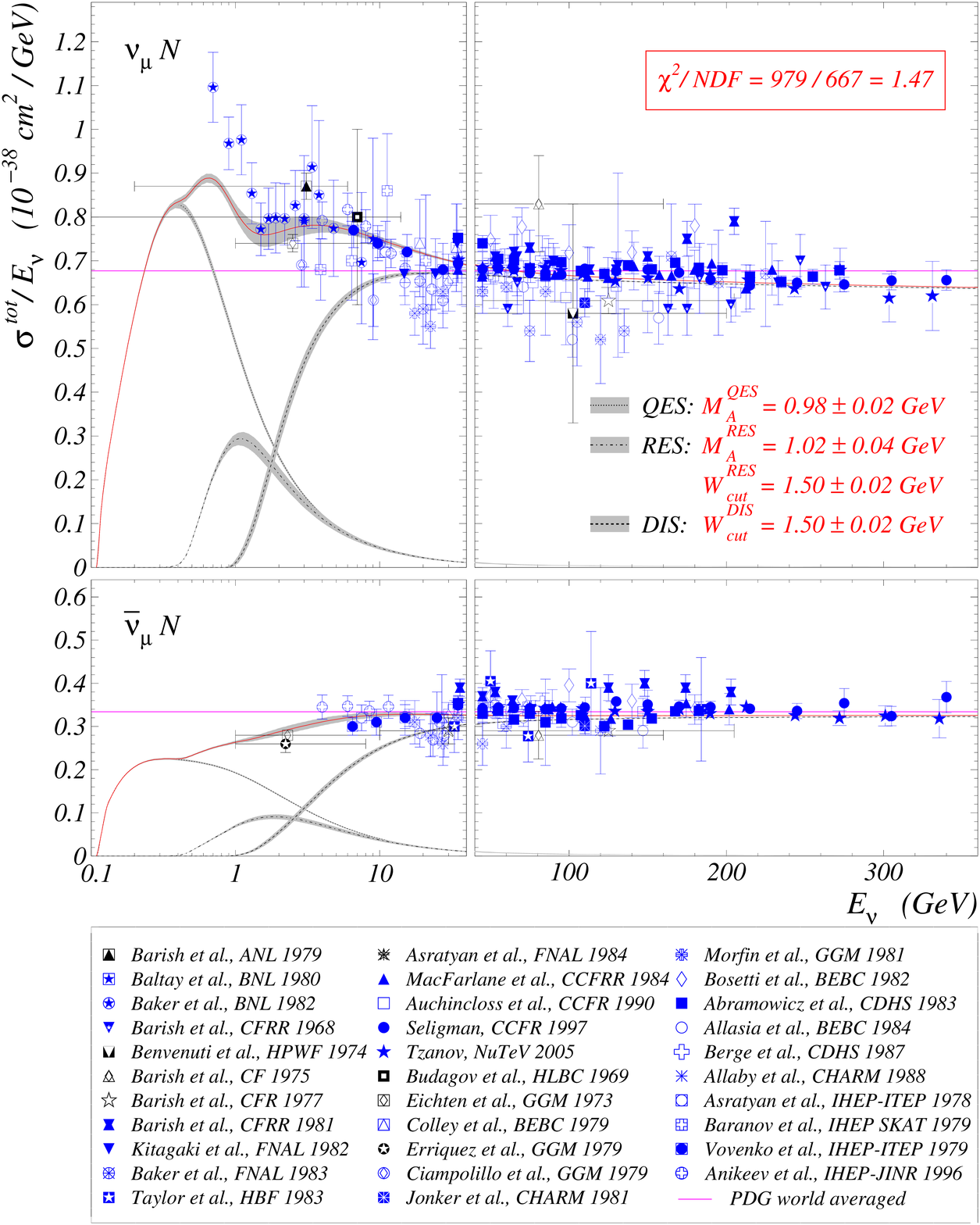}
\caption{Slopes of the total CC cross sections for $\nu_\mu$ and
         $\overline{\nu}_\mu$ scattering off an isoscalar nucleon
         measured by the experiments
         ANL~1979~\cite{Barish:79},
         BNL~1980~\cite{Baltay:80},
         BNL~1982~\cite{Baker:82},
         CFRR~1968~\cite{Barish:68},
         HPWF~1974~\cite{Benvenuti:74},
         CF~1975~\cite{Barish:75PRL},
         CFR~1977~\cite{Barish:77PRL},
         CFRR~1981~\cite{Barish:81},
         FNAL~1982~\cite{Kitagaki:82},
         FNAL~1983~\cite{Baker:83PRL},
         HBF~1983~\cite{Taylor:83},
         FNAL~1984~\cite{Asratyan:84b},
         CCFRR~1984~\cite{MacFarlane:84},
         CCFR~1990~\cite{Auchincloss:90},
         CCFR~1997~\cite{Seligman:97},
         NuTeV~2005~\cite{Tzanov:05},
         HLBC~1969~\cite{Budagov:69PLB},
         GGM~1973~\cite{Eichten:73PLBa},
         BEBC~1979~\cite{Colley:79},
         GGM~1979~\cite{Erriquez:79,Ciampolillo:79},
         CHARM~1981~\cite{Jonker:81},
         GGM~1981~\cite{Morfin:81},
         BEBC~1982~\cite{Bosetti:82},
         CDHS~1983~\cite{Abramowicz:83},
         BEBC~1984~\cite{Allasia:84},
         CDHS~1987~\cite{Berge:87},
         CHARM~1988~\cite{Allaby:88},
         IHEP-ITEP~1978~\cite{Asratyan:78},
         IHEP~SKAT~1979~\cite{Baranov:79},
         IHEP-ITEP~1979~\cite{Vovenko:79},
         and
         IHEP-JINR~1996~\cite{Anikeev:96}.
         The data points with horizontal error bars are for the
         slopes averaged over the wide energy ranges; these do
         not participate in the fit  and the corresponding energy binned data
         (included into the fit) are shown in
         Fig.~\protect\ref{Fig:SigmaSUM-I-FIT-16.02-f-6.02-2-0.80}.
         The curves and bands show the QES, RES, and DIS contributions and
         their sums calculated with the best-fitted values of the
         parameters depicted in the legend in top panel.
         The averaged values over all energies
         $(0.677\pm0.014)\times10^{-38}~\text{cm}^2/\text{GeV}$
         (for ${\nu}_{\mu}N$) and
         $(0.334\pm0.008)\times10^{-38}~\text{cm}^2/\text{GeV}$
         (for $\overline{\nu}_{\mu}N$) obtained by the Particle Data
         Group~\cite{Hagiwara:02} from the data by the experiments in
         Refs.~\cite{MacFarlane:84,Berge:87,Auchincloss:90,Seligman:97}
         are also shown for a comparison (straight lines).
\label{Fig:SlopeSUM-I-FIT-16.02-f-6.02-2-0.80}}

\end{figure}
\begin{figure}[htb]
\includegraphics[width=\linewidth]{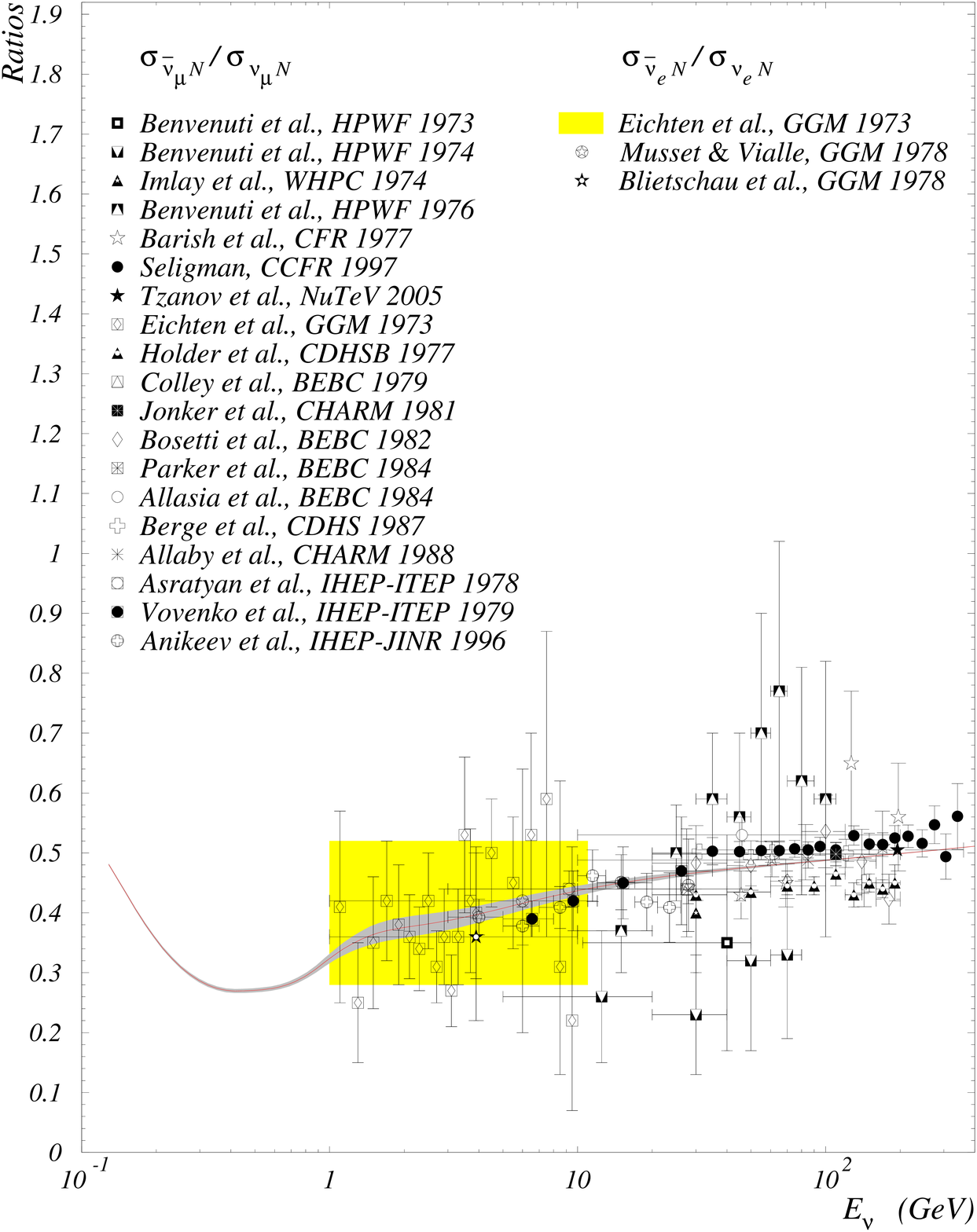}
\caption{The ratio $\sigma_{\overline{\nu}_{\mu}N}/\sigma_{\nu_{\mu}N}$
         for an isoscalar nucleon measured by the experiments
         HPWF~1973~\cite{Benvenuti:73},
         HPWF~1974~\cite{Benvenuti:74},
         WHPC~1974~\cite{Imlay:74},
         HPWF~1976~\cite{Benvenuti:76},
         CFR~1977~\cite{Barish:77PRL},
         CCFR~1997~\cite{Seligman:97},
         NuTeV~2005~\cite{Tzanov:05},
         GGM~1973~\cite{Eichten:73PLBa},
         CDHSB~1977~\cite{Holder:77},
         BEBC~1979~\cite{Colley:79},
         CHARM~1981~\cite{Jonker:81},
         BEBC~1982~\cite{Bosetti:82}
         (revised according to Ref.~\cite{Shotton:85}),
         BEBC~1984~\cite{Parker:84,Allasia:84},
         CDHS~1987~\cite{Berge:87},
         CHARM~1988~\cite{Allaby:86},
         IHEP-ITEP~1978~\cite{Asratyan:78},
         IHEP-ITEP~1979~\cite{Vovenko:79},
         and
         IHEP-JINR~1996~\cite{Anikeev:96}.
         The ratio $\sigma_{\overline{\nu}_eN}/\sigma_{\nu_eN}$
         reported in the three publications of the Gargamelle
         collaboration~\cite{Eichten:73PLBb,Musset:78,Blietschau:78}
         is also shown.
         The curve and band are calculated with the same values
         of the fitted parameters as in
         Fig.~\protect\ref{Fig:SlopeSUM-I-FIT-16.02-f-6.02-2-0.80}.
\label{Fig:RatioSUM-I-FIT-16.02-f-6.02-2-0.80}}
\end{figure}
\begin{figure}[htb]
\includegraphics[width=\linewidth]{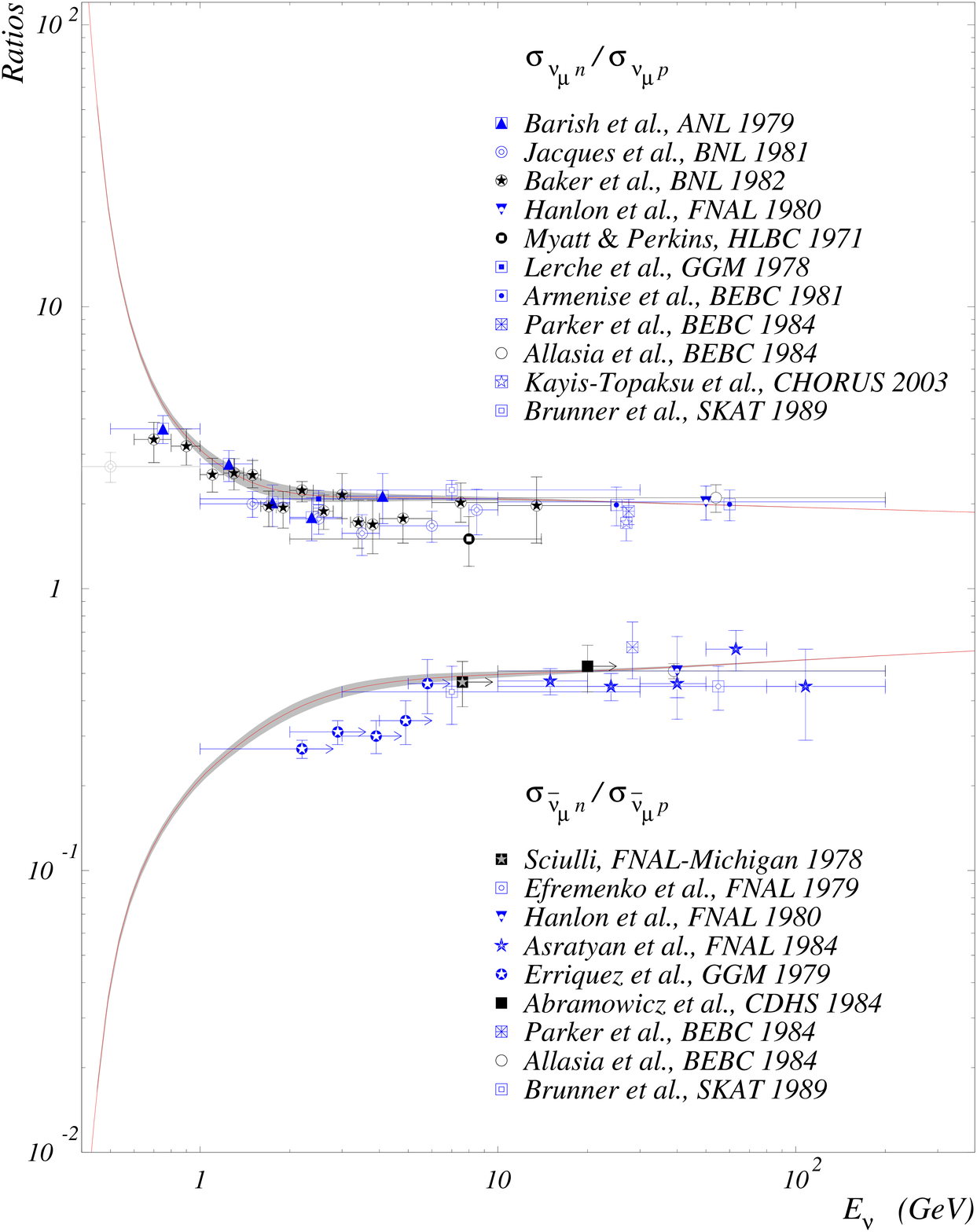}
\caption{The ratios $\sigma_{{\nu}n}/\sigma_{{\nu}p}$ and
         $\sigma_{\overline{\nu}n}/\sigma_{\overline{\nu}p}$
         measured by the experiments
         ANL~1979~\cite{Barish:79},
         BNL~1981~\cite{Jacques:81},
         BNL~1982~\cite{Baker:82},
         FNAL-Michigan~1978~\cite{Sciulli:78},
         FNAL~1979~\cite{Efremenko:79},
         FNAL~1980~\cite{Hanlon:80},
         FNAL~1984~\cite{Asratyan:84a},
         HLBC~1971~\cite{Myatt:71},
         GGM~1978~\cite{Lerche:78},
         GGM~1979~\cite{Erriquez:79},
         BEBC~1981~\cite{Armenise:81},
         CDHS~1984~\cite{Abramowicz:84},
         BEBC~1984~\cite{Parker:84,Allasia:84},
         CHORUS~2003~\cite{Kayis-Topaksu:02},
         and
         IHEP~SKAT~1989~\cite{Brunner:89ZPCa}.
         The data point of CDHS~1984 is recalculated in
         Ref.~\cite{Asratyan:84a} from the ratio 
         $\sigma_{\overline{\nu}p}/\sigma_{\overline{\nu}N}$.
         The curves and bands are calculated with the same values
         of the fitted parameters as in
         Fig.~\protect\ref{Fig:SlopeSUM-I-FIT-16.02-f-6.02-2-0.80}.
\label{Fig:RatioSUM-np-FIT-16.02-f-6.02-2-0.80}}
\end{figure}
\begin{figure}[htb]
\includegraphics[width=\linewidth]{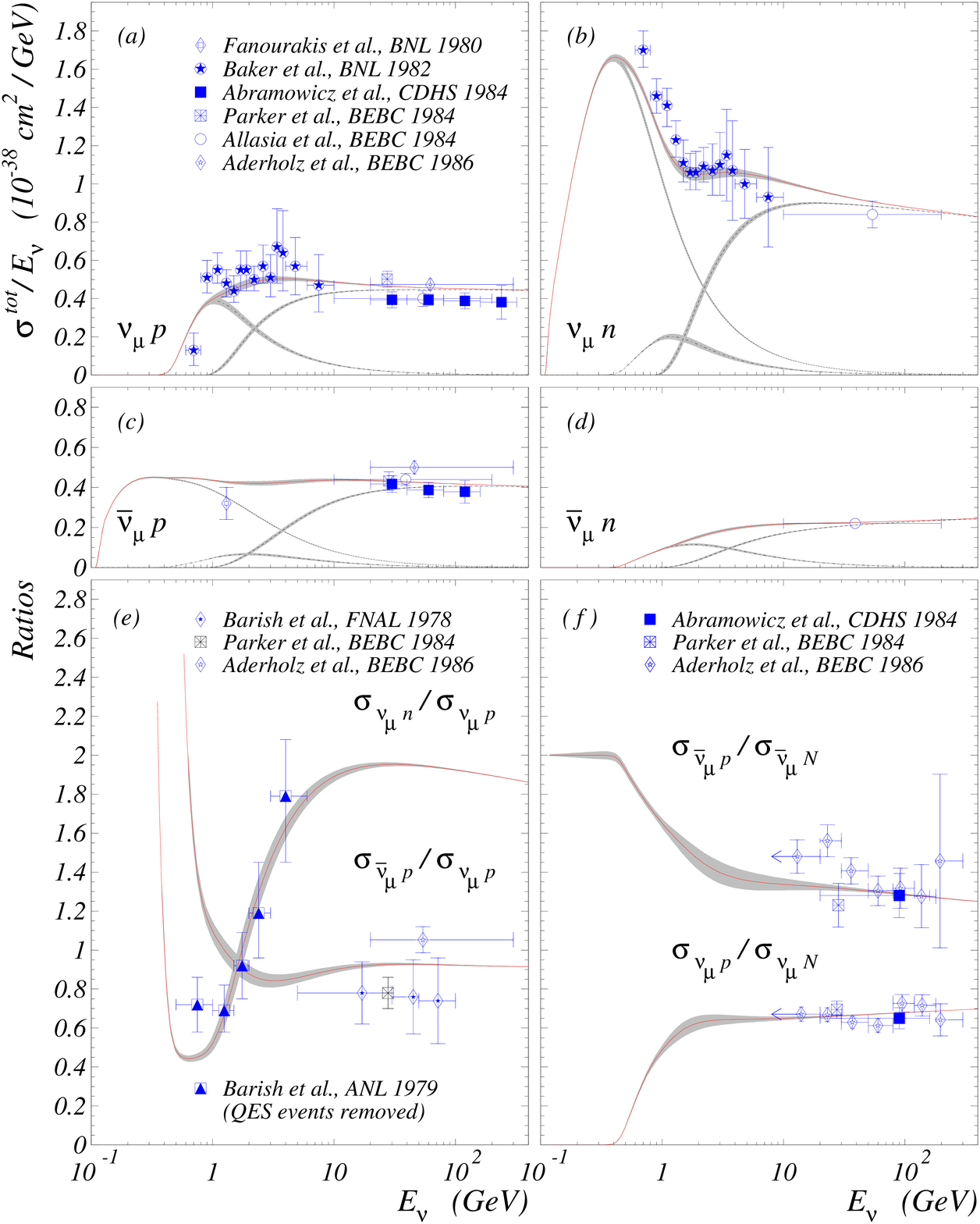}
\caption{(a), (b), (c), (d) --  the slopes of the ${\nu}_{\mu}p$,
         ${\nu}_{\mu}n$, $\overline{\nu}_{\mu}p$, and
         $\overline{\nu}_{\mu}n$ total CC cross sections measured
         by the experiments
         BNL~1980~\cite{Fanourakis:80},
         BNL~1982~\cite{Baker:82},
         CDHS~1984~\cite{Abramowicz:84},
         BEBC~1984~\cite{Parker:84,Allasia:84},
         and
         BEBC~1986~\cite{Aderholz:86};
         (e) -- the ratios 
         $\sigma_{{\nu}n}/\sigma_{\nu p}$ (with quasielastic events
         removed) and $\sigma_{\overline{\nu}p}/\sigma_{\nu p}$
         measured by the experiments
         ANL~1979~\cite{Barish:79},
         FNAL~1978~\cite{Barish:78},
         BEBC~1984~\cite{Parker:84},
         and
         BEBC~1986~\cite{Aderholz:86};
         (f) -- the ratios $\sigma_{{\nu}p}/\sigma_{{\nu}N}$ and
         $\sigma_{\overline{\nu}p}/\sigma_{\overline{\nu}N}$
         measured by the experiments
         CDHS~1984~\cite{Abramowicz:84},
         BEBC~1984~\cite{Parker:84},
         and
         BEBC~1986~\cite{Aderholz:86}.
         The curves and bands in all six panels are calculated with
         the same values of the fitted parameters as in
         Fig.~\protect\ref{Fig:SlopeSUM-I-FIT-16.02-f-6.02-2-0.80}.
\label{Fig:SlopeRatioSUM-npI-fit-16.02-f-6.02-2-0.80}}
\end{figure}

Figure~\ref{Fig:SigmaQES-Fit-16.02-f-6.02-2-0.80} shows the QES data from
Refs.~\cite{Kustom:69,Mann:73,Barish:77,Baker:81,Kitagaki:83,Suwonjandee:04,%
            Auerbach:02,Young:67,Budagov:69LNC,Bonetti:77,Armenise:79,Pohl:79,Allasia:90,%
            Makeev:81,Belikov:82,Belikov:85,Grabosch:88,Brunner:89ZPCb,Ammosov:92}
together with the B3 best fit to the \emph{full} set of the data satisfying the
criteria described in Sect.~\ref{Data set} ($\text{NDF}=670-3=667$).
The FNAL~1984 data points from Ref.~\cite{Asratyan:84b} (neon-hydrogen target)
are shown here only for a comparison.
They are not included into the fit since were obtained by a recalculation from
the DIS data (included into the fit, see
Fig.~\ref{Fig:SigmaSUM-I-FIT-16.02-f-6.02-2-0.80}) by using a prescription given
in Ref.~\cite{Asratyan:84b} and the errors for these points were estimated approximately.
In order to facilitate comparison, the data points for the experiments performed
with the nuclear targets different from $\text{D}_2$ and $\text{Ne-H}_2$ are converted to
a free nucleon target~\cite{Footnote_Nuclear_Effect}.
The nuclear effects for the
deuterium~\cite{Mann:73,Barish:77,Baker:81,Kitagaki:83,Allasia:90},
neon-hydrogen~\cite{Asratyan:84b} and
averaged iron data~\cite{Suwonjandee:04}
(shown in Fig.~\ref{Fig:SigmaQES-Fit-16.02-f-6.02-2-0.80} by filled rectangles)
were subtracted by the authors of the experiments.
The curves are calculated with $M_A^{\text{QES}}=0.98~\text{GeV}/c^2$,
the value obtained from the global B3 fit. The grey bands show the
standard deviation from the best-fit cross sections due to the error
of $0.02~\text{GeV}/c^2$ in determination of $M_A^{\text{QES}}$.
Note that the best-fit value of $M_A^{\text{QES}}$ is in agreement with
that obtained by a single-parameter fit to the QES data only,
$M_A^{\text{QES}}=0.94\pm0.04~\text{GeV}/c^2$.

The obtained value of $M_A^{\text{QES}}$, being lower, does not contradict
to the latest (still preliminary) result by the K2K experiment~\cite{Gran:05}
\[
M_A^{\text{QES}}(\text{K2K})=
1.06\pm0.03\,(\text{stat.})\pm0.14\,(\text{syst.})~\text{GeV}/c^2.
\]
It is however essentially below the value of $1.1~\text{GeV}/c^2$ used
in the recent atmospheric neutrino oscillation analysis of the
Super-Kamiokande~I experiment~\cite{Ashie:05}.

In Figs.~\ref{Fig:SigmaSUM-I-FIT-16.02-f-6.02-2-0.80} and
         \ref{Fig:SlopeSUM-I-FIT-16.02-f-6.02-2-0.80}
we collect the main subset of the experimental data on the total CC cross
sections and their slopes for an isoscalar nucleon
(hereafter denoted by $N$) from
Refs.~\cite{Barish:79,Baker:82,Benvenuti:74,Barish:75PRL,Barish:77PRL,%
            Asratyan:84b,Eichten:73PLBa,Eichten:73PLBa,Erriquez:79,%
            Ciampolillo:79,Budagov:69PLB,Baranov:79,Blietschau:78}
and    \cite{Barish:79,Baltay:80,Baker:82,Benvenuti:74,Barish:68,%
            Barish:75PRL,Barish:77PRL,Barish:81,MacFarlane:84,%
            Auchincloss:90,Kitagaki:82,Taylor:83,Baker:83PRL,Asratyan:84b,%
            Tzanov:05,Budagov:69PLB,Colley:79,Allasia:84,Eichten:73PLBa,%
            Erriquez:79,Ciampolillo:79,Morfin:81,Abramowicz:83,Berge:87,%
            Jonker:81,Allaby:88,Asratyan:78,Baranov:79,Vovenko:79,Anikeev:96},
respectively.
The majority of these data is included into the fit.
The curves and bands show the QES, RES, and DIS contributions
(Fig.~\ref{Fig:SlopeSUM-I-FIT-16.02-f-6.02-2-0.80}) and their sums (both figures)
calculated with the best-fitted values of $M_A^{\text{QES}}$, $M_A^{\text{RES}}$,
and $W_{\text{cut}}^{\text{RES}}=W_{\text{cut}}^{\text{DIS}}$ (the latter
equality is the restriction used in the B3 fit).

Figure~\ref{Fig:RatioSUM-I-FIT-16.02-f-6.02-2-0.80} accumulates the data on the
cross section ratios $\sigma_{\overline{\nu}_{\mu}N}/\sigma_{\nu_{\mu}N}$
and $\sigma_{\overline{\nu}_eN}/\sigma_{\nu_eN}$ (isoscalar target)
according to
Refs.~\cite{Benvenuti:73,Benvenuti:74,Benvenuti:76,Imlay:74,Barish:77PRL,%
            Seligman:97,Tzanov:05,Eichten:73PLBa,Eichten:73PLBb,Musset:78,%
            Blietschau:78,Colley:79,Bosetti:82,Allasia:84,Holder:77,%
            Berge:87,Jonker:81,Allaby:86,Asratyan:78,Vovenko:79,Anikeev:96}.
The paper~\cite{Blietschau:78} supersedes the earlier publications
of the Gargamelle Collaboration~\cite{Eichten:73PLBb} (shown by filled rectangle)
and \cite{Musset:78}.
The major part of these data is obtained from the cross sections measured in
the same experiments. The other, like the recent NuTeV result~\cite{Tzanov:05},
correspond to a wide energy range with no indication of the mean energy.
Due to these and similar reasons all these data are excluded from the analysis
and only shown here for a comparison with the result of the global B3 fit.
The cross section ratios $\sigma_{{\nu}n}/\sigma_{{\nu}p}$ and
$\sigma_{\overline{\nu}n}/\sigma_{\overline{\nu}p}$ from
Refs.~\cite{Barish:79,Jacques:81,Baker:82,Sciulli:78,Efremenko:79,Hanlon:80,%
            Asratyan:84a,Myatt:71,Armenise:81,Allasia:84,Abramowicz:84,Parker:84,%
            Lerche:78,Erriquez:79,Kayis-Topaksu:02,Brunner:89ZPCa}
are shown in Fig.~\ref{Fig:RatioSUM-np-FIT-16.02-f-6.02-2-0.80}.
The results of Refs.~\cite{Baker:82,Myatt:71,Sciulli:78,Allasia:84,Abramowicz:84}
are excluded from the fit due to the reasons mentioned in Sect.~\ref{Data set}.
The near-threshold point from Ref.~\cite{Jacques:81} is removed since its deviation
from the theoretical prediction is unphysically high.
Six panels of Fig.~\ref{Fig:SlopeRatioSUM-npI-fit-16.02-f-6.02-2-0.80} show the
data of different kinds from 
Refs.~\cite{Barish:79,Fanourakis:80,Baker:82,Barish:78,Abramowicz:84,Allasia:84,%
            Parker:84,Aderholz:86}.
Almost all data points participate in the analysis.
The curves and bands in
Figs.~\ref{Fig:RatioSUM-I-FIT-16.02-f-6.02-2-0.80},
      \ref{Fig:RatioSUM-np-FIT-16.02-f-6.02-2-0.80}, and
      \ref{Fig:SlopeRatioSUM-npI-fit-16.02-f-6.02-2-0.80}
are calculated with the parameters obtained from the global B3 fit
(see legend in Fig.~\protect\ref{Fig:SlopeSUM-I-FIT-16.02-f-6.02-2-0.80}).

\section{Summary}
  \label{Summary}

Our analysis of the world neutrino data on the QES and total CC cross sections
yields several thought-provoking conclusions.
As is seen from Tables \ref{Tab:FIT-16.02-f-6.02-2-0.80_1},
\ref{Tab:FIT-16.02-f-6.02-2-0.80_2}, and \ref{Tab:FIT-16.02-f-6.02-2-0.80_3},
in all variants of the fit there is a distinct minimum of $\chi^2$ for $M_A^{\text{QES}}$
around the ``canonical'' value of $1~\text{GeV}/c^2$ with deviations $\lesssim2\%$.
This is mainly an effect of the QES data subset whose exclusion from the analysis would
lead to an essential increase of $M_A^{\text{QES}}$ for all variants (for example, in the B3 and
A4 fits $M_A^{\text{QES}}$ becomes equal to $1.13\pm0.03$ and $1.17\pm0.03~\text{GeV}/c^2$,
respectively).

The situation with the $M_A^{\text{RES}}$ best-fit value is less definite:
in different variants of the fit it fluctuates from about $1.00$ to about
$1.15~\text{GeV}/c^2$. This spread comprises the BNL-2002 results for
$M_A^{\text{RES}}$~\cite{Furuno:03} obtained with different approaches
and does not contradict to the exact equality $M_A^{\text{RES}}=M_A^{\text{QES}}$.
However, the 3- and 4-parameter fits favour the case $M_A^{\text{RES}}>M_A^{\text{QES}}$.
Our ``favorable'' B3 variant of the fit yields the following values of the axial masses:
\[
M_A^{\text{QES}}=0.98\pm0.02~\text{GeV}/c^2
\]
and
\[
M_A^{\text{RES}}=1.02\pm0.04~\text{GeV}/c^2.
\]

The shape of the total and (all the more so) differential ${\nu}N$ and $\overline{\nu}N$
cross sections is very sensitive to the values of the cutoff parameters
$W_{\text{cut}}^{\text{RES}}$ and $W_{\text{cut}}^{\text{DIS}}$. From our analysis we have
to conclude that these parameters cannot be fine-tuned with the confidence level sufficient
for the current and future experiments for neutrino oscillations and related phenomena.
However, the most worth-while versions of the fit indicate that
$W_{\text{cut}}^{\text{RES}} \approx W_{\text{cut}}^{\text{DIS}}$
must be essentially above the value of $1.4~\text{GeV}$ approved in the
data processing of many accelerator and astrophysical neutrino experiments.
The outcome of the B3 fit is
\[
W_{\text{cut}}^{\text{RES}}=W_{\text{cut}}^{\text{DIS}}=1.50\pm0.02~\text{GeV}.
\]
Being considered deliberately, such a high value of the cutoff parameter for DIS
puts forward the difficult problem of a correct accounting for the reactions of
exclusive multi-hadron neutrinoproduction and coherent neutrino-nucleus scattering.

Finally we have to note that the above conclusions are only valid for the
theoretical models of the RES reactions, DIS structure functions and PDF,
as well as the approximations and simplifications (sometimes risky) adopted
in the present analysis.
Investigation of alternative models, a more accurate treatment of the nuclear effects,
and incorporation of additional experimental data is the matter of a forthcoming
work.

\acknowledgments

We are grateful to the Physics Department of Florence University and Theoretical
Department of KEK for warm hospitality during important stages of this work.
We thank to Yongguang Liang and Martin Tzanov for providing us with the
latest experimental data and relevant computer codes from JLab and NuTeV.
We also thank to Kaoru Hagiwara, Sergey Kruchinin, Kentarou Mawatari, Dmitry Naumov,
Roberto Petti, Gregory Vereshkov, and Hiroshi Yokoya for stimulating discussions.


\begin{thebibliography}{99}
\bibitem{Footnote_Lepton_Spin}
  As well as the final lepton spin, which is, however, not substantial
  to the present study.
\bibitem{Footnote_M_V}
  Considering that there are no solid reasons for choosing $M_V=0.84~\text{GeV}/c^2$,
  the parameter $M_V$ is a natural candidate to be on the list of free parameters
  in a refined likelihood analysis involving the experimental data on the
  single-pion neutrinoproduction.
\bibitem{Footnote_Tau}
  Not the case for the $\nu_{\tau}N$ and $\overline{\nu}_{\tau}N$ DIS processes.
\bibitem{Footnote_Nuclear_Effect}
   Actually the fit comprehends the primary data for heavy targets while
   the nuclear effects are included into the calculated QES cross sections.
\end{thebibliography}
\end{document}